# Atomically resolved electron reflectivity at a metal/semiconductor interface

Ding-Ming Huang,[1,2,3,*] Jian-Huan Wang,[1,2,3] Jie-Yin Zhang,[4] Yuan Yao,[3] H. Q. Xu,[1,2,†] and Jian-Jun Zhang[3,4,5,‡]

[1]*Beijing Academy of Quantum Information Sciences, Beijing 100193, China*

[2]*Beijing Key Laboratory of Quantum Devices and School of Electronics, Peking University, Beijing 100871, China*

[3]*Beijing National Laboratory for Condensed Matter Physics and Institute of Physics, Chinese Academy of Sciences, Beijing 100190, China*

[4]*Songshan Lake Materials Laboratory, Dongguan 523808, China*

[5]*Hefei National Laboratory, Hefei 230088, China*

(Dated: Oct. 9, 2025)

**Abstract**

An atomically flat interface is achieved between face-centered cubic Al and diamond lattice Ge via molecular beam epitaxy (MBE). Based on the measurements of scanning tunneling microscopy (STM), we demonstrate an atomically resolved lateral periodic change of the electron reflectivity at the Al/Ge interface. The variation of electron reflectivity is up to 24% in lateral 2 nm. We speculate that the change of reflectivity results from the local electronic states at the Al/Ge interface. This phenomenon provides an atomically non-destructive method for detecting the buried interfacial states in hetero-structures by STM.

Epitaxial metal thin films on semiconductor substrates exhibit exotic physical properties [1-7]. Due to vertical thickness quantum confinement and lateral lattice modulations by substrates, the electronic properties of these metal thin films are strongly related to the crystalline quality and lattice structures at the interfaces. Epitaxially grown Pb and Al thin films on Si substrate are two well-studied examples, where quantum size effect induced modulations of electron-phonon coupling (EPC) [1,2] and superconductivity [3,4], the emergence of type-II superconductivity [4-6] and Mott transition [7] were experimentally observed. Non-destructive characterizations of interfacial structures and properties in these materials are essential. Over the past two decades, studies have demonstrated that the interfacial lattice structures between Si substrate and epitaxial metal (Pb, In, Cd and Al) thin films can be observed through scanning tunneling microscopy (STM) surface imaging [8-14], even if the films are up to a few dozen atomic layers thick [8]. The STM visualizing of the buried interfaces relies on two types of mechanisms. First, in the condition of a metal film grown on a reconstructed substrate, the visualization of the buried interface results from both lateral change of film thickness and local electron diffuse scattering at the interfacial vacancies, where the diffuse scattering may also be related to local electron-phonon scattering

(EPS) [8,11]. Second, in the condition of an adiabatic interface [12], where interfacial vacancies are absent, the interface visualization is achieved via measuring the lateral phase shift in the electron reflective scattering, which is sensitively modulated by the interfacial atomic arrangements [12-14].

Here, we report the epitaxial growth of single crystalline Al(111) films on a Ge(111) substrate, as well as experimental study of the interfacial properties of the as-grown samples by STM and scanning tunneling spectroscopy (STS) measurements. A distinct Moiré pattern of the interfacial Al/Ge lattice is observed on the Al surface by STM, which is persistent for film thicknesses ranging from 1 nm to 25 nm (4~100 ML). Based on the STM and STS measurements, we demonstrate the presence of an adiabatic and reflective-phase-homogeneous Al/Ge interface. Thus, the observed Moiré pattern in our system is independent of local electron diffuse scattering and reflective phase shift induced by interfacial atomic arrangements, in contrast to previous studies. By considering the Al film as a Fabry-Perot interferometer, we have introduced a model to describe the electron transmission in the STM measurements and shown that such model qualitatively reproduces our experimental observations. These results allow us to attribute the Moiré pattern to the atomically resolved lateral changes of electron reflectivity at the perfect Al/Ge interfaces.

To obtain a single crystalline Al film on Ge, a 30 nm thick Ge buffer is firstly grown on Ge (111) using molecular beam epitaxy (MBE) to achieve a high-quality Ge surface. The sample is then *in-situ* transferred to another MBE chamber for Al deposition. The Al film is deposited at 40 $^o$C with a growth rate of 1.0 Å/s. Figure 1a shows a cross-sectional high-angle-annular-dark-field (HAADF) scanning transmission electron microscope (STEM) image of the Al/Ge interface. It is seen that the face-centered cubic (FCC) Al perfectly sits on the diamond structured Ge with an atomic-level flat interface. The STM measurements are performed at a temperature of 10 K. The STM image on a 2 nm thick Al film is shown in Fig. 1b. Here, a 2 nm periodic Moiré pattern is visible and the black dashed-line diamond marks a unit cell. Atomic resolution STM image and the corresponding Fast-Fourier-Transform (FFT) are shown in the insets of Fig. 1b. Reciprocal lattice points of the Al and the buried Ge are marked by $q_1$ and $q_2$, respectively, while $q_3$ and $q_4$ are reciprocal points of the Moiré period with $q_3=2q_2-q_1$ and $q_4=q_1-q_2$. The lattice constants on (111) facets of bulk Al and Ge are 2.86 Å and 4.00 Å, respectively. A 7-Al-atom lattice on Al(111) matches to a 5-Ge-atom lattice on Ge(111) (mismatch is only about 0.1%), forming a commensurate 2 nm periodicity, as confirmed by the HAADF STEM (Fig. 1a) and the STM (Fig. 1b). Figure 1c schematically shows the possible atomic arrangement (top-view) at the Al/Ge interface with a unit cell marked by dashed diamond. A simulation of the Moiré pattern based on this lattice configuration can be found in the supporting information [15]. Temperature dependent low-energy electron diffraction (LEED), STEM and STM studies demonstrate a negligible strain-related distortion on Al surface lattice, and the Moiré pattern is related to vertical coherent electronic states in the Al film [15].

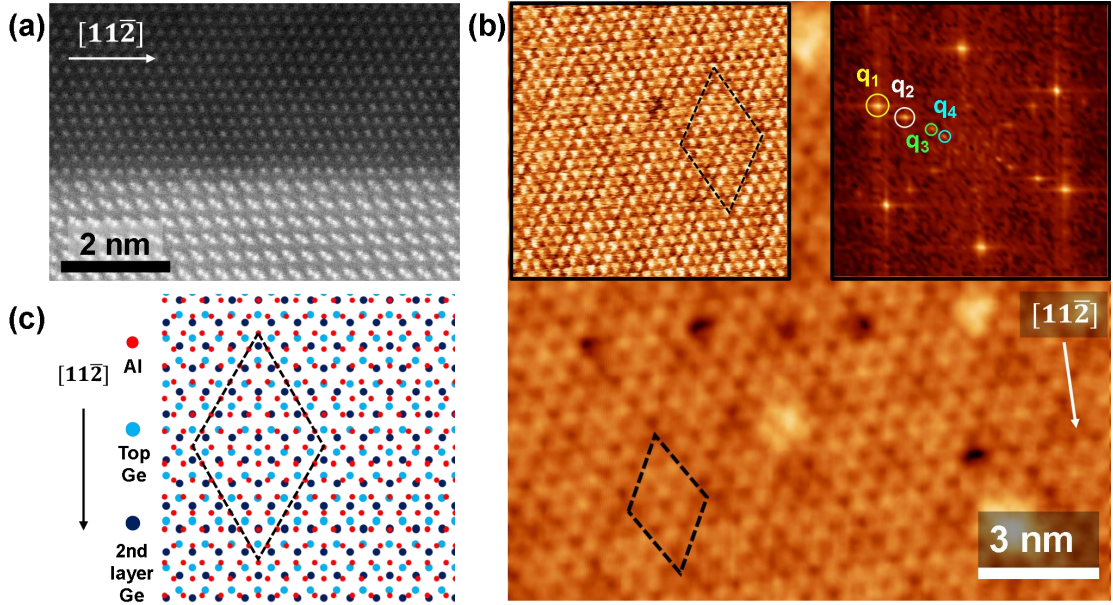

**Fig. 1.** Epitaxial Al film on Ge(111). (a) Atomic resolution cross-sectional STEM HAADF image of the Al/Ge interface. (b) STM tomographic image on a 2 nm thick Al/Ge film (taken at sample bias of $V_s$ = -60 mV and tunneling current of $I_t$ = -318 pA). A 2 nm periodic interfacial Al/Ge supercell is marked by a dashed diamond. The top-left inset shows the atomic resolution STM image of the top-layer Al ($V_s$ = -70 mV, $I_t$ = -1 nA). The top-right inset shows the FFT image. The reciprocal lattice points of Al and Ge are marked by $q_1$ and $q_2$, respectively. $q_3$ and $q_4$ are derived points of $q_3=2q_2-q_1$ and $q_4=q_1-q_2$, respectively. (c) Schematic of the in-plane atomic arrangement at the Al/Ge interface. The grey dashed diamond marks a supercell.

The quantum well (QW) states in Al films are studied by STS measurements via a standard lock-in technique. It is noted that a pre-treated blunt tip is used to take the STS spectra [15], ensuring a high energetic precision. Figure 2a shows an averaged differential conductance (dI/dV) spectrum taken on a 10 nm thick Al film. Several peaks are observed, which are attributed to the QW states in the Al film. Figure 2c shows a set of dI/dV spectra obtained from different positions within a Moiré pattern supercell, as denoted by colored cross markers in Fig. 2b. We see that the dI/dV peaks at different positions appear almost at the same sample bias of 104.5 ± 3.5 mV, indicating an atomically flat Al/Ge interface and negligible lateral variation of the interfacial reflective phase [8,12-14]. As a comparison, the dI/dV spectrum (black curve in Fig. 2c) taken at an atomic disorder site, most probably a Ge dopant [15], shows a large QW energy shift of -25.5 meV or a significant phase change. These results imply that the STM visualizing of the buried Al/Ge interface is not caused by a local reflective phase shift or a film thickness variation [8,12-14].

Now, we analyze the EPS at different positions of the Moiré pattern and demonstrate that the STM visualized pattern is not related to a local electron diffuse scattering either. The EPS is characterized by the electron-phonon spectral function (Eliashberg function) $\alpha^2F(\omega)$, which can be obtained from the $d^2I/dV^2$ spectra [2,11,16]. The QW states have significant influence on the strength of EPS [2], leading to modulations of the intensity in $d^2I/dV^2$ spectra [15]. Despite the influence from QW states, the lateral change in the local $d^2I/dV^2$ spectra is negligible. Figure 2d shows the $d^2I/dV^2$ spectra obtained by numerical differentiation of the $dI/dV$ spectra in Fig. 2c. It is evident that the peak position of 32 mV (matching the energy of phonons in Al [17~19]) as well as the peak values are both independent of the lateral position. This lateral homogeneity of the $d^2I/dV^2$ spectra indicates that there is no observable variation in the local EPS and thus, the STM observed pattern is not related to a local electron diffuse scattering at the interface [11].

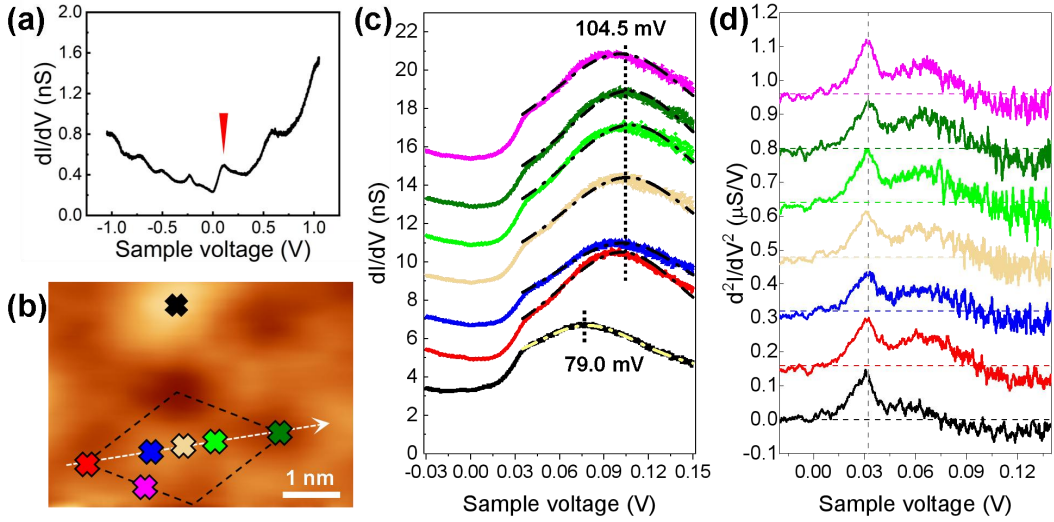

**Fig. 2.** STS spectra on the Moiré pattern. (a) A $dI/dV$ spectrum on a 10 nm thick Al film (stabilized at $V_s$ = -450 mV, $I_t$ = -125 pA, bias modulation of $V_{mod}$ = 25 mV). The oscillation peaks in the curve arise from tunneling through QW states. The red arrow points to the QW peak studied in Fig. 2c. (b) STM image of a 2 nm periodic Moiré pattern supercell ($V_s$ = -50 mV, $I_t$ = -65 pA). Colored crosses mark the positions for STS measurements. (c) Local $dI/dV$ spectra ($V_s$ = -50 mV, $I_t$ = -165 pA, $V_{mod}$ = 5 mV) taken at the positions marked by colored crosses in (b). These curves are shifted vertically by 2 nS for clarity. The dot-dashed curves are the fittings to the measurements. There is a peak shift of -25.5 mV in the black curve, which is taken at the position containing a Ge dopant. (d). The corresponding $d^2I/dV^2$ spectra obtained by numerical differentiation of the $dI/dV$ spectra in (c). These curves are shifted vertically by 0.16 μS/V for clarity. The peak at 32 mV is attributed to EPS in the Al film. The peak values are independent of the lateral position.

We observe that the width of the QW peaks shown in Fig. 2c depends on the lateral position. For example, the peak measured at the position marked by the red cross in Fig. 2b (the red curve in Fig. 2c) exhibits a sharper profile compared to the width of the peak taken at the position marked by the blue cross in Fig. 2b (the blue curve in Fig. 2c). Similar phenomenon has also been observed in STS measurements of Pb/Si systems [20, 21]. Since the coherence of these QW states depends on the boundary conditions, the electronic states in Al/Ge interface have considerable influence on the STM results. To understand the lateral variation of the peak width, a one-dimensional model is introduced to describe the electron transmission in an STM measurement. Figure 3a schematically shows the band diagram and the coherent transmission of an electron through the Al film. Due to strong Fermi-level pinning at an Al/Ge interface, the interfacial Fermi-level is pinned near the valence band edge of Ge [15, 22, 23]. Metal-induced gap states (MIGS) are formed on Ge side with a spatial depth of several atomic layers [15, 22, 24, 25]. The Al film is treated as a Fabry-Perot interferometer with two reflective interfaces, i.e., the Al/Ge interface and the vacuum/Al interface. An electron can experience multiple reflections in the Al cavity, and the transmissivity of a coherent electron is a function of the electron energy of $E$. In an ideal interferometer (with no dissipation within the cavity), the transmissivity of a coherent electron is given by the Fabry-Perot spectral function [26-28]:

$$T = \frac{1}{1+\frac{4f^2}{\pi^2}\sin^2\left[k_z(E)L+\frac{\Phi}{2}\right]}. \qquad (1)$$

Here, $f$ is the interferometer finesse and is a function of the electron reflectivity at the two interfaces. In the Al cavity, $f$ depends on the interfacial states. Typically, a higher local density of states results in a higher tunneling rate across the Al/Ge boundary, i. e. a lower reflectivity and a smaller $f$. $k_z(E)$ is the electron wavevector along the growth direction, and $L$ is the thickness of the Al film. $\Phi$ is the sum of the reflective phase shift at the Al/Ge interface and the Al surface, and the value is independent of the lateral position (as discussed in Fig. 2c). The transmissivity of $T$ exhibits an oscillatory behavior with increasing $k_z(E)$. The peak value of $T$ appears when the electron wavevector matches the QW coherent condition of the Al film, which is given by

$$2k_z(E_n)L + \Phi = 2n\pi, \qquad (2)$$

where $n$ is the quantum number. The wavevector difference between the neighboring QW states of $n+1$ and $n$, denoted as $\Delta k_z$, can be calculated as $\Delta k_z = \pi/L$. We have $\Delta k_z \approx 0.03$ Å$^{-1}$ in an Al film of $L$=10 nm. Since the $\Delta k_z$ is significantly smaller than the size of Al Brillouin zone of 1.55 Å$^{-1}$, the derivative of $k_z(E)$ to $E$, denotes as $\delta k_z/\delta E$, is approximately to

$$\frac{\delta k_z}{\delta E}|_{E_n} = \frac{\pi}{L\Delta E_n}, \qquad (3)$$

where $\Delta E_n$ is the energy difference between the QW states of $n$ and $n+1$.

In realistic conditions of STM measurements on an Al/Ge film, the differential conductance dI/dV can be written as a sum of two parts, coherent tunneling part and incoherent tunneling part, respectively. Bring Eq. (3) to Eq. (1), the differential conductance near the Fermi-Level has a relation of the following [27, 28]:

$$\frac{dI}{dV} \propto A(E) \cdot \frac{1}{1+\frac{4f^2}{\pi^2}\sin^2\left[\frac{\pi}{\Delta E_{nF}}(E-E_{nF})\right]} + B(E) \quad \text{for } (E_F<E<E_{nF+1}), \qquad (4)$$

where the coefficient $A(E)$ is a smooth function of $E$ and is dependent on the tip properties and the tip setup with respect to the Al surface. $B(E)$ is the incoherent tunneling term describing the contributions of electrons dissipated before reaching the Al/Ge interface and is also normally a smooth function of $E$. $E_F$ is the Fermi-Level, $E_{nF}$ and $E_{nF+1}$ are the energies of the first and second QW states above the Fermi-Level, and $\Delta E_{nF}$ is the energy difference between the QW states just below and above the Fermi-Level. Equation (4) indicates that a peak value of the differential conductance appears at the electron energy of $E_{nF}$, corresponding to a QW state. The full width at half maximum (FWHM) of the peak is dominantly determined by the coherent tunneling term and is a function of $f$, expressed as $\Delta E_{\text{FWHM}} \approx 2\sin^{-1}(\pi/2f) \cdot \Delta E_{nF}/\pi$. Therefore, a larger $f$ corresponds to a sharper peak in the dI/dV spectra.

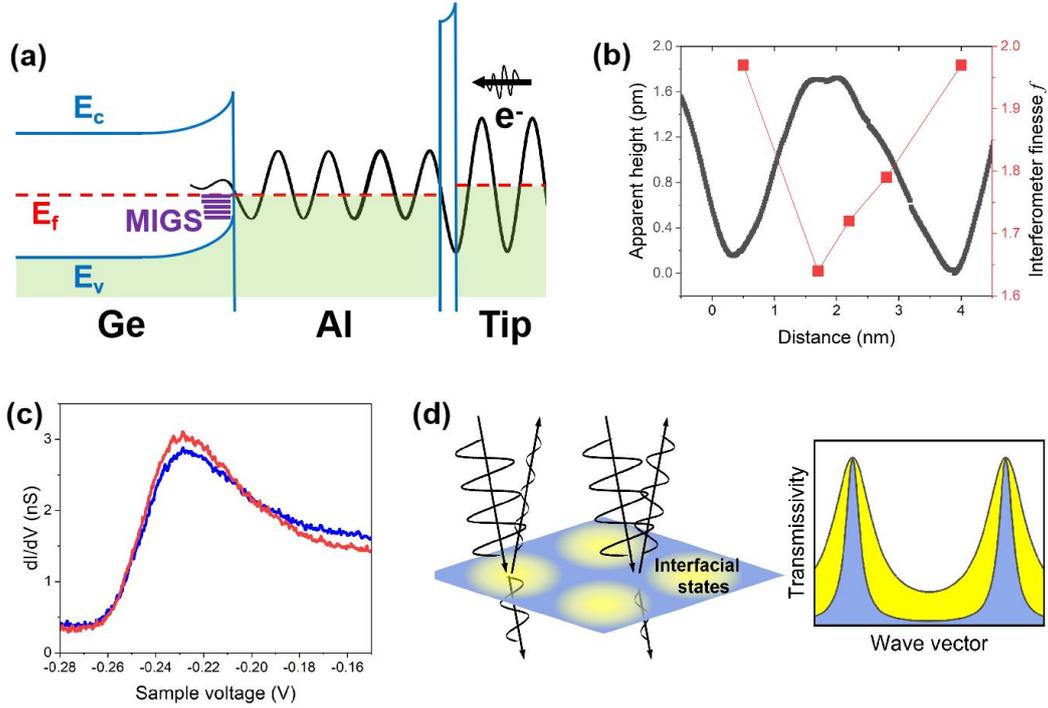

**Fig. 3.** Lateral variation of the interfacial electron reflectivity. (a) Schematic of the band diagram of an Al/Ge film and the transmission process of a coherent electron in STM measurements. A coherent electron can experience multiple reflections in the Al film, and the transmissivity from tip to the MIGS depends on the interfacial reflectivity. (b) The STM apparent height and fitted values of $f$ acquired along the white dashed line in Fig. 2b. The value of $f$ decreases when the apparent height increases. (c) The corresponding dI/dV spectra ($V_s$ = -50 mV, $I_t$ = -135 pA, $V_{\text{mod}}$ = 3 mV) of the QW state at -227.5 meV measured at two lateral positions marked by the red and blue crosses in Fig. 2b, respectively. The spectrum obtained at a position with a lower apparent height (red) shows a sharper peak. (d) Schematic of the lateral changes in electron transmissivity. Interfacial states result in a periodic lateral change in electronic reflectivity, leading to variations in the transmissivity of electrons through the Al cavity. The contour lines of transmissivity at different lateral positions are filled with the corresponding colors.

By fitting the dI/dV spectra to Eq. (4), the values of $f$ for each curve in Fig. 2c are determined.

It is worth noting that the data obtained at sample biases below +35 mV are excluded from the fittings, because the EPS results in an "inelastic tunneling dip" in this bias range [28, 29]. Considering that the dI/dV spectra shown in Fig. 2c were taken in the same tip setup and the fitting to the data was performed in a small energy range of 35 < $E$ < 150 meV, A($E$) was treated as a constant. The QW energy difference of $\Delta E_{nF}$ in a 10 nm thick Al film is 332 meV, which is obtained by measuring the peak positions in the dI/dV spectrum. Fitting results are shown in dashed lines in Fig. 2c, which can qualitatively reproduce our experimental results. The fitted value of B($E$) for each curve is at least one order of magnitude smaller than the coherent term (except the black colored curve obtained on a defected site), indicating that the coherent tunneling is dominant in STM measurements. Figure 3b shows the STM apparent heights and the fitted values of $f$ obtained at different positions along the white dashed line in Fig. 2b. It is clear that a higher $f$ corresponds to a lower apparent height. According to Eq. (1), a higher $f$ leads to a lower transmissivity of coherent electrons. The suppression of electron transmissivity would result in a lower local apparent height on the atomically flat Al surface in the measurements where the tip approaches to the sample to sustain the current in constant-current scanning mode. The relation between $f$ and the apparent height is in good agreement with our model. Figure 3c shows the dI/dV spectra of the QW state at -227.5 meV measured at two lateral positions marked by the red and blue crosses in Fig. 2b. Here, it is seen that the width variation of the spectra peak is consistent with the observations in Fig. 2c, i.e., a sharper peak corresponds to a measuring position with a lower STM apparent height. This result indicates that the position dependence of $f$ is almost unchanged in the electron energy range of several hundreds meV. Therefore, the Moiré pattern in Fig. 1b (taken at sample bias of $V_s$ = -60 mV) and Fig. 2b ($V_s$ = -50 mV) reflects the spatial variation of $f$ for electrons near the Fermi-Level.

We argue that the lateral changes in the interferometer finesse $f$ are caused by the variations of electron reflectivity at the Al/Ge interface. In the realistic vacuum/Al/Ge interferometer with dissipation, the $f$ is given by [26, 27]

$$f = \frac{\pi(R_1 R_2)^{1/4} \cdot e^{-L/2\lambda}}{1 - \sqrt{R_1 R_2} \cdot e^{-L/\lambda}}, \tag{5}$$

where $R_1$ and $R_2$ are the electron reflectivity on the Al/Ge interface and Al/vacuum interface, respectively. The $R_2$ on the single crystalline Al(111) is considered as a constant. The $\lambda$ is the electron mean free path, which describes the effects of diffuse scatterings from lattice defects or phonons. The presence of defects is related to an energy shift in the peaks in dI/dV spectra (black curve in Fig. 2c). All the dI/dV spectra used for analyzing $f$ were obtained from defect-free regions, which have a constant dI/dV peak position of 104.5 mV. On the other hand, as shown in Fig. 2d, the strength of EPS is independent of the lateral position. Therefore, the lateral change in $\lambda$ is negligible as the scatterings from defects and phonons are laterally homogeneous, and the variations of $f$ are attributed to changes of $R_1$. Bring the fitted values of $f$ to Eq. (5), the peak-to-peak lateral change of the electron reflectivity is calculated as $\Delta R_1/R_1$ = 24%. The periodic changes in the reflectivity is attributed to the local electronic states in the Al/Ge interface. Figure

3d schematically shows the lateral changes of electron transmissivity. A higher local reflectivity results in a lower transmissivity for electrons mismatched to the QW coherent condition in the Al film, leading to a sharper width of the peak in transmissivity. The local dI/dV spectra on samples with various thicknesses are studied [15]. The position dependence of the QW peaks in these experiments is all qualitatively consistent with Fig. 3d, indicating that the physical model matches the experimental results of various thicknesses.

Spatial dI/dV maps of various sample biases are studied, the details are shown in supporting information [15]. The contrast of the Moiré pattern in the dI/dV maps changes at different sample biases. The lowest contrast appears at the bias corresponding to the QW energy of $E_n$, indicating that the spatial variation of differential conductance is suppressed when the electron is perfectly matched to the QW coherent condition. This result is also in good agreement with our physical model of Eq. (4).

In conclusion, we have shown that the interfacial atomic structures in an Al/Ge epitaxial film can be observed by STM imaging. Different from the previously studied metal/semiconductor thin films, the electronic scattering at the atomically flat Al/Ge interface exhibits lateral uniformity in both the reflective phase and electron-phonon scattering strength. We have introduced a physical model based on a Fabry-Perot interferometer to describe the electron transmission in the STM measurements, which successfully explains the experimental observations. From this model, we conclude that the visualization of the buried interface results from the lateral change in interfacial electron reflectivity induced by interfacial states. The lateral modulations of electron reflectivity may be general in various hetero-interfaces. We believe that our observations have potential applications for non-destructive detection of adiabatic and atomic-level flat interfaces, such as interfaces in Van der Waals stacking of layered materials or incoherent heterostructures with exotic interlayer couplings.

Acknowledgements

We thank Kai Chang, Yun Liu for discussions. We thank Yixin Chu, Binxiao Fu and Ting Yip Chan for the MBE maintenance and writing improvement. We thank Xue-Tao Zhu, Tian Qian, Zhi-Yu Tao, En-Ling Wang and Ren-Jie Zhang for the characterization of the Al surface lattice and the energy band. This work was supported by the NSFC (Nos. 62225407, 92165207, 92165208, 12304207, 12304101, 12304100), and the Innovation Program for Quantum Science and Technology (No. 2021ZD0302300).

\* huangdm@baqis.ac.cn

† hqxu@pku.edu.cn

‡ jjzhang@iphy.ac.cn

Ding-Ming Huang and Jian-Huan Wang contributed equally to this paper.

Reference:

[1] Y. F. Zhang, J. F. Jia, T. Z. Han, Z. Tang, Q. T. Shen, Y. Guo, Z. Q. Qiu, and Q. K. Xue, Phys. Rev. Lett. **95**, 096802 (2005).

[2] M. Schackert, T. Märkl, J. Jandke, M. Hölzer, S. Ostanin, E. K. U. Gross, A. Ernst, and W. Wulfhekel, Phys. Rev. Lett. **114**, 047002 (2015).

[3] Y. Guo, Y. F. Zhang, X. Y. Bao, T. Z. Han, Z. Tang, L. X. Zhang, W. G. Zhu, E. G. Wang, Q. Niu, Z. Q. Qiu, J. F. Jia, Z. X. Zhao, and Q. K. Xue, Science **306**, 1915 (2004).

[4] W. M. J. van Weerdenburg, A. Kamlapure, E. H. Fyhn, X. Huang, N. P. E. van Mullekom, M. Steinbrecher, P. Krogstrup, J. Linder, and A. A. Khajetoorians, Sci. Adv. **9**, eadf5500 (2023).

[5] T. Zhang, P. Cheng, W. J. Li, Y. J. Sun, G. Wang, X. G. Zhu, K. He, L. Wang, X. Ma, X. Chen, Y. Wang, Y. Liu, H. Q. Lin, J. F. Jia, and Q. K. Xue, Nat. Phys. **6**, 104 (2010).

[6] Y. X. Ning, C. L. Song, Y. L. Wang, X. Chen, J. F. Jia, Q. K. Xue, and X. C. Ma, J. Phys.: Condens. Matter **22** 065701 (2010).

[7] I. B. Altfeder, X. Liang, T. Yamada, D. M. Chen, and V. Narayanamurti, Phys. Rev. Lett. **92**, 226404 (2004).

[8] I. B. Altfeder, D. M. Chen, and K. A. Matveev, Phys. Rev. Lett. **80**, 4895 (1998).

[9] Y. Jiang, K. H. Wu, Z. Tang, Ph. Ebert, and E. G. Wang, Phys. Rev. B **76**, 035409 (2007).

[10] M. L. Tao, H. F. Xiao, K. Sun, Y. B. Tu, H. K. Yuan, Z. H. Xiong, J. Z. Wang, and Q. K. Xue, Phys. Rev. B **96**, 125410 (2017).

[11] I. B. Altfeder, K. A. Matveev, and A. A. Voevodin, Phys. Rev. Lett. **109**, 166402 (2012).

[12] I. B. Altfeder, V. Narayanamurti, and D. M. Chen, Phys. Rev. Lett. **88**, 206801 (2002).

[13] W. B. Jian, W. B. Su, C. S. Chang, and T.T. Tsong, Phys. Rev. Lett. **90**, 196603 (2003).

[14] Y. Jiang, J. D. Guo, Ph. Ebert, E. G. Wang, and K. H. Wu, Phys. Rev. B **81**, 033405 (2010).

[15] See supporting information.

[16] E. Minamitani, N. Takagi, R. Arafune, T. Frederiksen, T. Komeda, H. Ueba, and S. Watanabe, Prog. Surf. Sci. **93**, 131 (2018).

[17] S. Y. Savrasov, D. Y. Savrasov, and O. K. Andersen, Phys. Rev. Lett. **72**, 372 (1994).

[18] G. G. Rusina, S. V. Eremeev, S. D. Borisova, I. Yu. Sklyadneva, and E. V. Chulkov, Phys. Rev. B **71**, 245401 (2005).

[19] R. Bauer, A. Schmid, P. Pavone, and D. Strauch, Phys. Rev. B **57**, 11276 (1998).

[20] S. M. Lu, M. C. Yang, W. B. Su, C. L. Jiang, T. Hsu, C. S. Chang, and T. T. Tsong, Phys. Rev. B **75**, 113402 (2007).

[21] Y. S. Fu, S. H. Ji, T. Zhang, X. Chen, J. F. Jia, Q. K. Xue, and X. C. Ma, Chin. Phys. Lett. **27**, 066804 (2010).

[22] T. Nishimura, K. Kita, and A. Toriumi, Appl. Phys. Lett. **91**, 123123 (2007).

[23] A. Dimoulas, P. Tsipas, A. Sotiropoulos, and E. K. Evangelou, Appl. Phys. Lett. **89**, 252110 (2006).

[24] S. Ciraci, A. Baratoff, and I. P. Batra, Phys. Rev. B **43**, 7046 (1991).

[25] V. Heine. Phys. Rev. **138**, A1689 (1965).

[26] M. Fox, Quantum Optics: An Introduction (Oxford University Press, USA, 2006).

[27] J. J. Paggel, T. Miller, and T. C. Chiang, Science **283**, 1709 (1999).

[28] K. Wang, X. Zhang, M. M. T. Loy, T. C. Chiang, and X. Xiao, Phys. Rev. Lett. **102**, 076801 (2009).

[29] X. Wu, C. Xu, K. Wang, and X. Xiao, Phys. Rev. B **92**, 035434 (2015).

# SUPPORTING INFORMATION

**SUPPORTING INFORMATION 1**:
**A simulation of Moiré pattern based on the interfacial lattice.**

In Fig. SI 1a, on the left part we schematically plot the possible lattice configuration of the Al/Ge interface, and on the right part the corresponding simulated Moiré pattern is shown. The unit cell is marked by dashed diamond. The FFT image of this lattice is shown in Fig. SI 1b, where q1 and q2 are reciprocal points corresponding to individual Al and Ge lattices, respectively. The reciprocal points of q3 and q4 show the highest intensities within the Ge Brillouin Zone, corresponding to the points of q3=2q2-q1 and q4=q1-q2, respectively, which perfectly match the measured points in **inset of Fig. 1b**. By filtering the points of q3 and q4, we obtain the Moiré period as shown in the schematic of the right part of Fig. SI 1a, which matches the experimental patterns.

It is necessary to point out that this simulation provides only a qualitative result. Since the Moiré pattern is correlated to the interfacial states, in-depth considerations of interfacial coupling will be necessary for more accurate simulations.

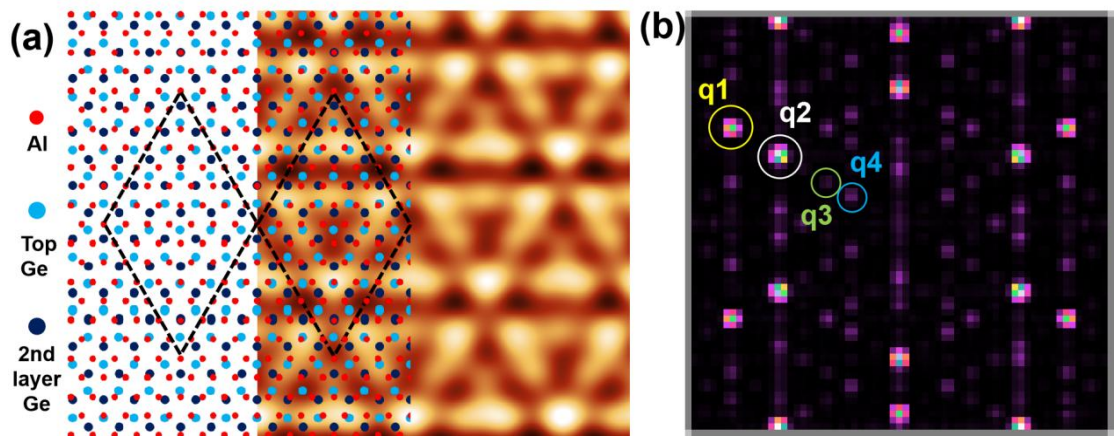

**Fig. SI 1.** Simulation of the Moiré pattern. (a) Schematic of a possible interfacial lattice and the corresponding Moiré pattern. Unit cells are marked by dashed diamonds. (b) The FFT image of the lattice in (a). The reciprocal lattice points of Al and Ge are marked by q1 and q2, respectively. q3 and q4 are derived points of q3=2q2-q1 and q4=q1-q2, respectively.

**SUPPORTING INFORMATION 2:**

**The correlation between the Moiré pattern and the electronic coherent states**

Decoherence of electron can be achieved by increasing the temperature, as the electron mean free path λ decreases due to the enhancement of electron-phonon scattering (EPS). Fig. SI 2a and 2c show the FFT images of STM morphology on a 25 nm thick Al/Ge film obtained at 10 K and room temperature (RT), respectively. The reciprocal points of the Al/Ge interfacial lattice are shown at 10 K and vanish at RT. The dI/dV spectra obtained at 10K and RT are shown in Fig. SI 2b and 2d, respectively. The differential conductance is a smooth function of sample bias at RT, as the decoherence of electrons causes the vanishing of the QW peaks. The presence of the Moiré pattern coincides with the presence of QW states, further indicating that the Moiré pattern is caused by the coherent electronic states.

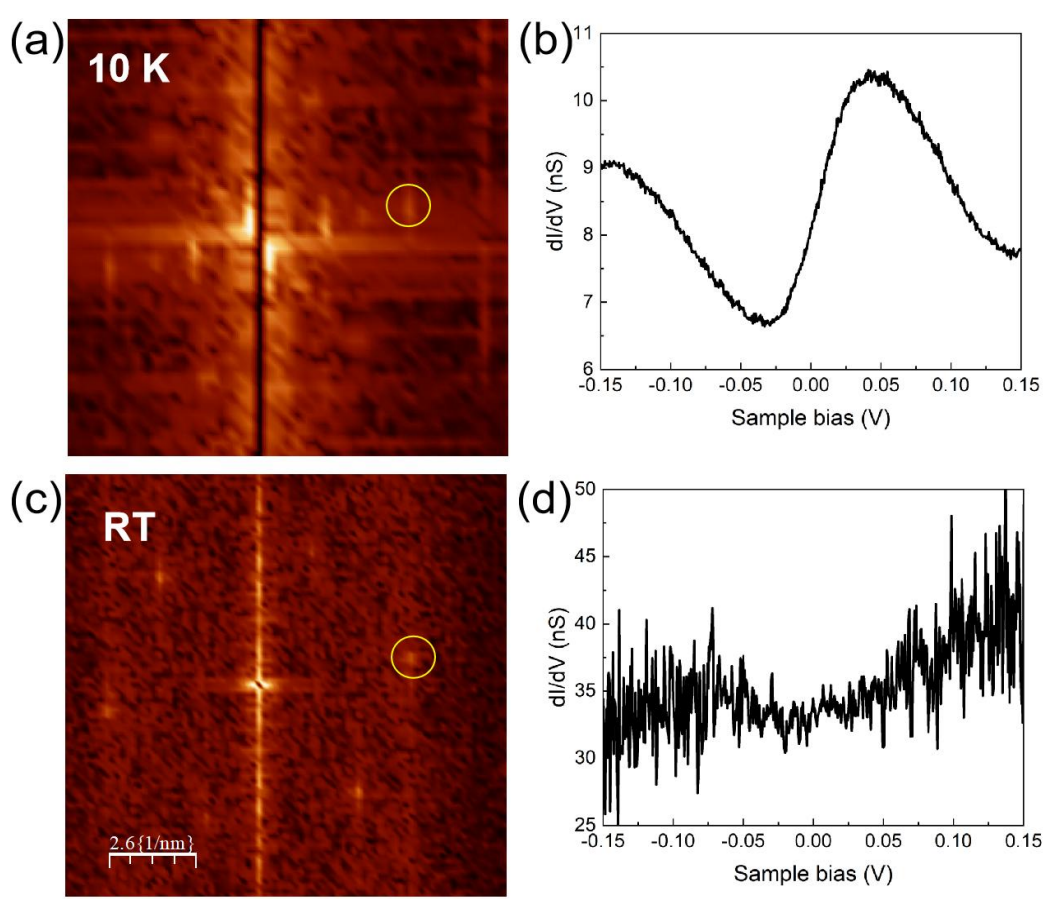

**Fig. SI 2.** Decoherence of reflective electron with increasing of temperature in a 25 nm thick Al/Ge film. (a) FFT image of the surface morphology at 10 K. A reciprocal point of Al lattice is marked by yellow circle. (b) dI/dV spectrum on surface at 10 K (-50 mV, -345 pA and 28 mV modulation). (c) FFT image of the surface morphology at RT. (d) dI/dV spectrum on surface at RT (-50 mV, -1.14 nA and 5 mV modulation). A reciprocal point of Al lattice is marked by yellow circle.

# SUPPORTING INFORMATION 3：
## Characterizations of the distortion of Al lattice

The surface lattice of a 10 nm thick Al/Ge film is characterized by an in-situ Low-Energy Electron Diffraction (LEED). Figure SI 3a and Fig. SI 3b are LEED images taken at RT and 35 K, respectively. To achieve the highest imaging contrast, the energy of induced electrons is set at 132 eV, and only 3 reciprocal points of the Al surface lattice are visualized. The contrast of LEED images enhances with the decreasing of temperature, but the LEED pattern is invariant. The absence of Moiré points indicates that the reconstruction of the Al surface lattice is negligible (at least smaller than the resolution of LEED), and the Moiré pattern in STM measurements is dominant by local electronic property. The STM image on a stacking fault is shown in Fig. SI 3c, where the disorder morphology is only visible within 3 nm from the stacking fault. The Moiré pattern near the stacking fault (beyond 3 nm) remains the same as the pattern on defect-free region. While, the stacking fault induced distortion at least influence the lattice within lateral 15 nm, which can be observed by the height variations in the STM image (Fig. SI 3d and 3e). Thus, the Moiré pattern is independent of lattice distortion.

Furthermore, we introduce an FFT analysis of scanning-transmission-electron-microscope (STEM) images to study the lattice structure. A cross-sectional STEM image of the Al/Ge film is shown in Fig. SI 3f, and the FFT image of the area within the dashed square is presented in Fig. SI 3g. The FFT image only shows the reciprocal points of Al lattice, indicating the absence of Moiré period in the lattice structure.

In summary, the Moiré pattern in Al/Ge film is independent of lattice reconstruction and distortion.

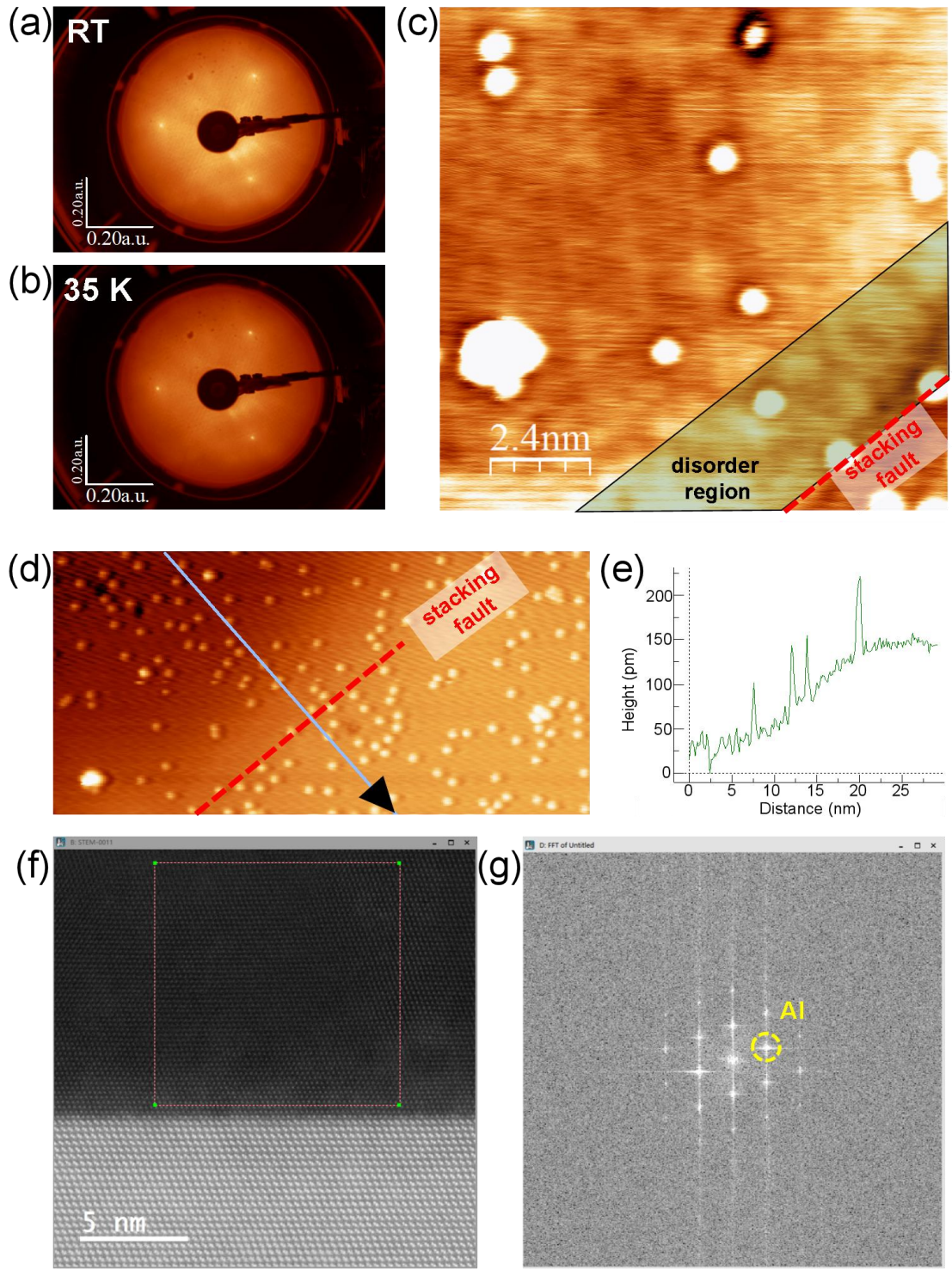

**Fig. SI 3.** Surface lattice and defect induced distortion on a 10 nm thick Al/Ge film. (a) and (b) LEED images at RT and 35 K, respectively. (c) Moiré pattern near a stacking fault. (d) STM morphology on a stacking fault. (e) Profile of height taken from the grey line in (d). (f) STEM High-angle-annular-dark-field image of the Al/Ge film. (g) FFT image of the selected area in (f). Only the reciprocal points (marked by dashed ring) of Al lattice are observed.

**SUPPORTING INFORMATION 4:**
**Pre-treatment of STM tip for dI/dV studies.**

Achieving atomic resolution image on low Miller index metal surfaces (such as Al(111)) is always challenging due to the "plain" surface DOS and small lattice constant, especially for an s-wave STM tip with "plain" DOS at Fermi-level (Phys. Rev. Lett. **65**, 448 (1990)). The atomic resolution image in the inset of **Fig. 1b of manuscript** is obtained by a "Ge decorated tip", which is obtained by picking up a Ge cluster (or a Ge atom). However, this tip cannot be utilized in STS measurements due to the unpredictable DOS of the attached cluster. In order to acquire STS spectra with high energy precision, tips for STS measurements are pre-treated on an Au(111) substrate until a standard dI/dV spectrum of Au surface is achieved (an example is shown in Fig. SI 4). The calibrated tips present an s-wave feature and typically have a reduced spatial resolution.

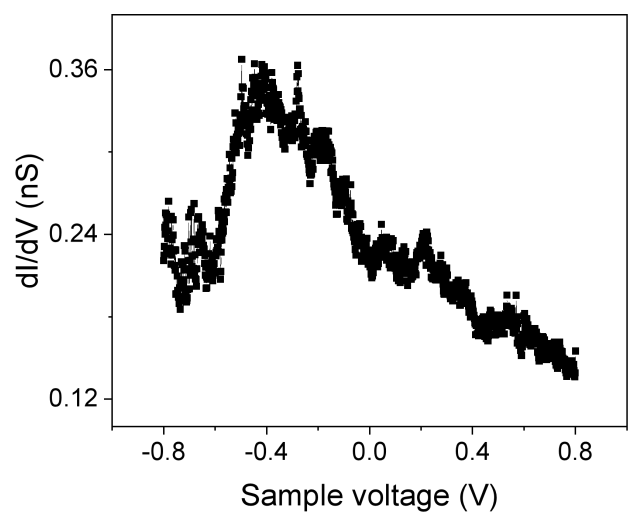

**Fig. SI 4.** dI/dV spectrum taken on Au(111) surface after the calibration of tip's DOS.

# SUPPORTING INFORMATION 5:
**Point defects in the Al/Ge films.**

We speculate that the defect site (black marked) in **Fig. 2b** is caused by Ge dopant. Below is a detailed discussion of our assumption.

Figure SI 5a shows an atomic resolution STM image of 3 types of point defects and the corresponding height profiles, from top to bottom are "bright spot", surface vacancy and adatom, respectively. The apparent height of "black spot" defects in **Fig. 1b of manuscript** are the same as that of vacancies. Therefore, we can determine that the "dark spots" in Fig. 1b are vacancies. Figure SI 5b shows the height profile of the point defect in **Fig. 2b of manuscript**, and the apparent height is close to that of "bright spot" in Fig. SI 5a, indicating that these defects are the same. According to the atomic resolution image of "bright spot" in Fig. SI 5a, where the most top atoms are clearly visible, we can determine that this defect is located under the Al surface. Figure SI 5c to f schematically show 4 types of point defects that may form under the Al surface, which are interfacial Ge vacancy, interfacial Al dopant, Al vacancy and Ge dopant, respectively. The dI/dV spectrum on the "bright spot" defect is shown in **Fig. 2c of manuscript**, where the QW peak shift of black curve (towards negative bias) indicates a higher effective thickness or an increase in effective energy barrier (PRB **81**, 033405 (2010)). We speculate that the Ge dopant (Fig. SI 5f) is the only reason that could lead to the STM results we obtained.

The interfacial Ge vacancy (Fig. SI 5c) may cause a decreased apparent height at surface due to the strain induced by the missing atom. This defect breaks three Ge-Ge valence bonds, and the dangling bonds may attract an Al atom, forming an Al dopant (Fig. SI 5d). The interfacial Al dopant result in a slight increase in effective thickness. However, similar to Ga dopant on Si substrate (PRB **81**, 033405 (2010)), an Al dopant on Ge may also induce a screening of the interfacial barrier, leading an additional QW peak shift towards positive bias. Al vacancy (Fig. SI 5e) may result in a slight decrease of the effective thickness, leading a QW peak shift to positive bias and a decrease in apparent height. Different to the 3 types of defects being mentioned above, Ge dopants (Fig. SI 5f) may lead to STM results we obtained. The relatively larger size of Ge atoms may result in an increased apparent height and a corresponding QW shift. Furthermore, the fitting of dI/dV (black curve in **Fig. 2c**) demonstrate a significant incoherent tunneling at the defect site. The fitting value of B(E) in Eq. (4) is comparable to the value of A(E) (A/B = 1.1). This result indicates an enhanced incoherent scattering, which attribute to scatterings from the Ge dopant.

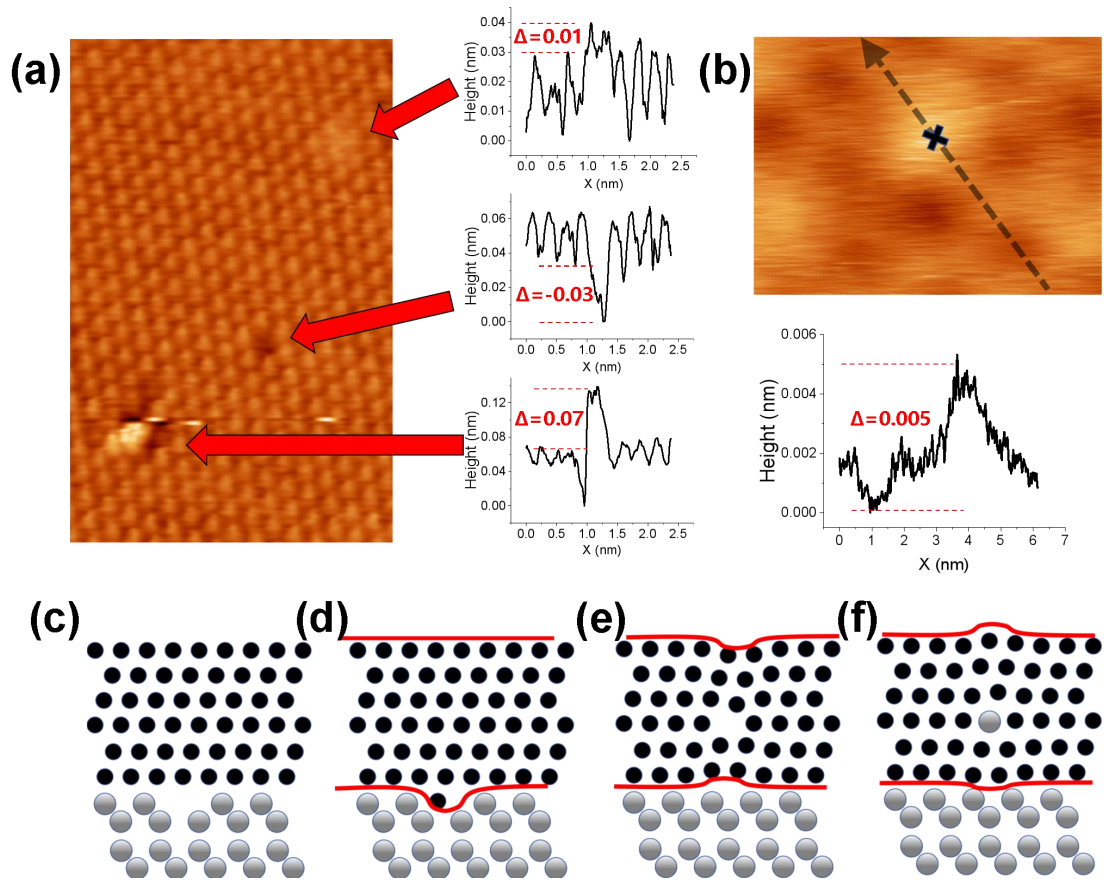

**Fig. SI 5.** Point defects in the material. (a) Atomic resolution STM image of point defects and the corresponding height profile. From top to bottom are "bright spot", surface vacancy and adatom, respectively. (b) The STM image of **Fig. 2c** in manuscript and the height profile. The apparent height of +5 pm proves that the defect is not a surface adatom. (c) to (f) Schematic of 4 types of point defects that can occur in Al films. (c) interfacial Ge vacancy, (d) interfacial Al dopant, (e) bulk Al vacancy and (f) Ge dopant, respectively. The red curves schematically illustrate the lateral variation of effective thickness caused by corresponding defects.

## SUPPORTING INFORMATION 6:
Details of local $d^2I/dV^2$ spectra.

The QW states significantly influence the electron-phonon-interaction (EPI) in our Al/Ge films, leading to modulations of intensity in $d^2I/dV^2$ spectra, as also reported in Phys. Rev. Lett. **114**, 047002 (2015). In our experiments, the $d^2I/dV^2$ peak positions at positive and negative biases are perfectly matched, but the intensity varies. Despite the asymmetric feature of $d^2I/dV^2$ spectra, unchanged $d^2I/dV^2$ peaks at different positions indicate that the Moiré pattern is independent of EPI. Below are the details.

Local $d^2I/dV^2$ spectra on a 10 nm thick Al/Ge sample, measured with bias modulation of 2 mV, are shown in Fig. SI 6a, red and blue curves are measured on positions corresponding to colored marks in Fig. 2b. The peaks at ± 32 mV and ± 23 mV (triangle marked) result from the EPI in Al [17-19]. The peak value at positive bias is approximately double of that at negative bias, this asymmetry persists across various measurements and independent of bias modulation or tunneling current. We attribute this asymmetry to the result of QW-states-modulated-electron-phonon-interaction (PRL **114**, 047002 (2015)). The corresponding local dI/dV spectra are shown in Fig. SI 6b (displayed in corresponding colors), where the QW peak is close to the Fermi-level and results in a steep slope at positive bias, supporting our hypothesis. In order to reduce the influence from QW states, we studied the $d^2I/dV^2$ spectra on a 3 nm Al/Ge film, where the Fermi-level is approximately centered between two QW peaks. Fig. SI 6c shows the $d^2I/dV^2$ data, and the measuring positions are marked in Fig. SI 6d with corresponding colors. The inset of Fig. SI 6d shows the dI/dV spectrum taken on this surface, which presents QW peaks at -380 mV and +440 mV, respectively. Fig. SI 6c shows a slight enhancement of $d^2I/dV^2$ at negative bias. This asymmetry is still attribute to the QW enhancement, as the QW peak at negative bias is slightly closer to the Fermi-level. These $d^2I/dV^2$ results demonstrate that the asymmetry is influenced by the positions of the QW peaks

Despite the asymmetric feature, the difference between $d^2I/dV^2$ spectra in Fig. SI 6a is negligible, indicating homogeneous EPI across different positions. Furthermore, we would like to point out that **if we have missed any of the ultra-fine EPI features, this EPI cannot result in a strong Moiré pattern.** In summary, the Moiré pattern is independent of EPI.

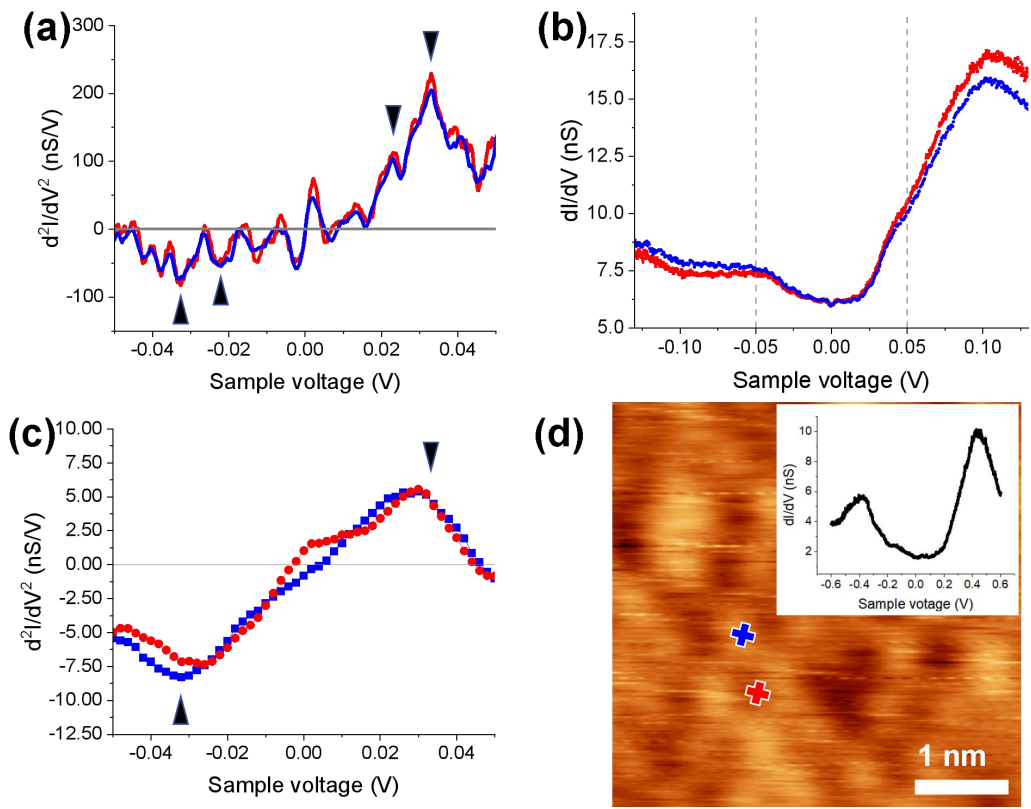

**Fig. SI 6.** Local $d^2I/dV^2$ spectra on the Moiré pattern. (a) and (b) Local $d^2I/dV^2$ spectra and $dI/dV$ spectra on a 10 nm Al/Ge film. These spectra are measured at the corresponding positions described in **Fig. 2**. Triangles in (a) mark the $d^2I/dV^2$ peaks at ± 32 mV and ± 23 mV, resulting from electron-phonon interactions in Al. (c) Local $d^2I/dV^2$ spectra obtained on a 3 nm film. The triangles mark the voltage value of ± 32 mV. (d) STM image of the Moiré pattern on a 3 nm Al/Ge film. Inset shows the $dI/dV$ spectrum taken on the surface. The colored crosses mark the measuring positions of the $d^2I/dV^2$ curves in (c).

**SUPPORTING INFORMATION 7:**
**Fermi-level pinning at Al/Ge interface.**

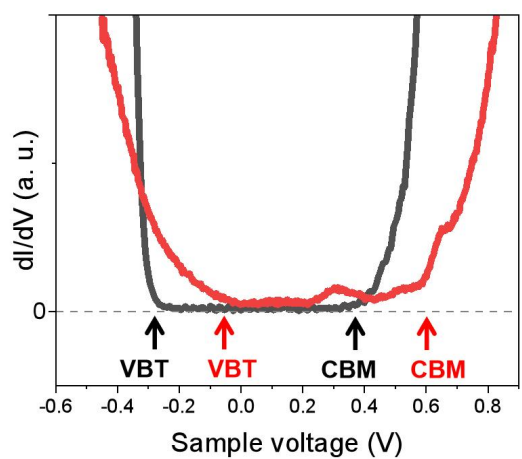

**Fig. SI 7.** dI/dV spectra on an epitaxial Ge film and a 0.8-ML- Al covered Ge(111) surfaces. The results on intrinsic Ge(111) and Al/Ge(111) are shown in black and red curves, respectively. The valence-band-top (VBT) and conduction-band-minimum (CBM) are marked by colored arrows.

The band edges of Ge are characterized by STS. As shown in Fig. SI 7, the black curve represents the dI/dV spectrum taken from an intrinsic epitaxial Ge(111) film, while the red curve indicates spectrum from a 0.8-monolayer-Al-covered Ge(111). The deposition of Al layer was performed at the same condition described in the paper main text. Previous studies showed that the valence-band-top (VBT) of clean Ge(111) depends on doping (PRB **71**, 125316 (2005)), i. e. the Fermi-level is not pinned at Ge(111) surface. Here, we see that the VBT of our Ge(111) layers (black curve) sits at -0.28 eV below the Fermi-level. In comparison, the VBT with Al coverage (red curve) has shifted by +0.22 eV towards Fermi-level. The VBT on our Al/Ge(111) sample is almost the same as the value observed on other metal/Ge layers (Appl. Phys. Lett. **91**, 123123 (2007)), indicating that the Fermi-level pinning occurs on this commensurate interface. These dI/dV spectra indicate that the Fermi-level pinning results from the coverage of Al, which is related to metal-induced-gap-states.

**SUPPORTING INFORMATION 8:**
**dI/dV results on various film thicknesses.**

The dI/dV spectra on Al/Ge films show dramatic changes with increase of thicknesses, which result from the QW peak shift. An example of this QW shift is shown in Fig. SI 8a, which is obtained on a 10 nm thick Al/Ge sample. The thickness dependence of the dI/dV spectra on Al/Ge samples was studied and shown in Fig. SI 8c to h. It is worth noting that the changes in STM resolution in Fig. SI 8d and e are due to the random tip shapes after pre-treatments (details can be found in **supporting information 4**). Fig. SI 8c shows the local dI/dV spectra (obtained with a same tip height) and the Moiré pattern on a 2 nm Al/Ge sample. The dI/dV measuring positions are marked by crosses with corresponding colors, and the red site presents a lower apparent height than the blue one. A unit cell of the Moiré pattern is marked by a dashed diamond. The red curve presents lower dI/dV values near the Fermi-level than the blue one, i. e. a sharper QW peak. These dI/dV spectra agree with the physical model of Eq. (1). Fig. SI 8d to h show the STM results obtained on 3 nm, 4 nm, 6 nm, 7 nm and 8 nm samples, respectively. In order to minimize the error from thermal drift, the dI/dV spectra of these thicknesses (including the results in manuscript main text) are obtained under "constant current mode", where the tip height is recalibrated by current set-point before each individual measurement. A schematic of the "constant current mode" measurement is shown in Fig. SI 8b. The tip approaches the sample at positions with lower electron transmissivity, keeping the dI/dV values near Fermi-level constant and result in a higher QW peak. All the dI/dV data from Fig. SI 8d to h show higher QW peaks at low apparent height sites (red), demonstrating that the Fabry-Perot interferometer model matches the experimental results under various thicknesses.

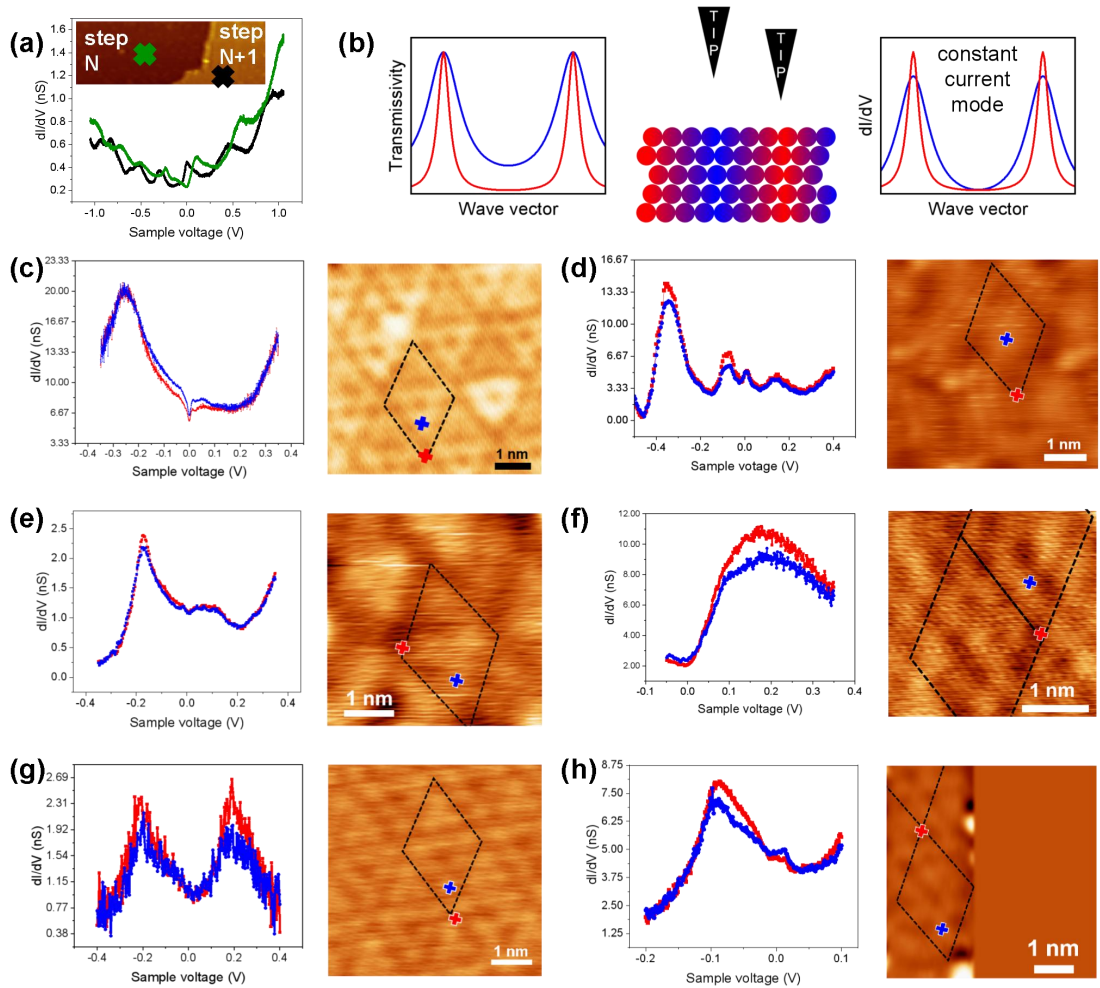

**Fig. SI 8.** Local dI/dV spectra on Al/Ge films with various thicknesses. (a) dI/dV spectra obtained from neighbored atomic steps on a 10 nm Al film. QW peaks present an energy shift of about -110 meV with the increasing of a single atomic step. Inset: STM image shows the atomic step. The dI/dV measurement sites are marked by colored crosses. (b) Schematic of dI/dV measurements under "constant current mode". When STM tip approaches the sample positions with lower electron transmissivity, it results in a higher peak value. (c) Local dI/dV spectra and STM image obtained on a 2 nm thick Al/Ge film. (d) to (h) Local dI/dV spectra and STM images obtained on 3 nm, 4 nm, 6 nm, 7 nm and 8 nm thick Al/Ge samples, respectively. These dI/dV spectra are measured under "constant current mode". The dI/dV measurement sites and the unit cells of Moiré pattern in STM images from (c) to (h) are marked by colored crosses and dashed diamonds, respectively. Blue dI/dV curves from (c) to (h) are obtained at positions (blue marks) with a higher apparent height.

**SUPPORTING INFORMATION 9:**
**Studies of spatial dI/dV maps.**

We utilize the self-correlation and FFT methods to study the Moiré period in dI/dV maps, these are the common methods applied in analysis of in-plane periods (PRB **89**, 235115 (2014) ).

Figure SI 9a to c present the dI/dV maps at sample biases of +203 mV, +102 mV and +82 mV, respectively. The corresponding self-correlation and FFT images of the dI/dV maps are shown in the left and right insets, respectively. The dI/dV spectrum taken at the mapping region is shown in Fig. SI 9d, and a QW peak sits at +79±3 mV. Cross-sectional profiles along the marked lines in self-correlation images are shown in Fig. SI 9e. The periodic oscillation of intensity is suppressed when the sample bias reduces to +82 mV, indicating a decrease in periodic correlation. Cross-sectional profiles along the marked lines in FFT images are shown in Fig. SI 9f. The intensity peaks at 3.6 nm$^{-1}$ correspond to the reciprocal point of q4 in **Fig. 1b**. The peak value reduces with decreasing of sample bias from +203 mV to +82 mV, indicating a suppression in periodic discrepancy of dI/dV. According to the physical model of Eq. (1) in the paper's main text, the lateral variation of the electronic transmissivity disappears when the electron wavevector perfectly matches to the coherent condition in a Fabry-Perot cavity. Therefore, the suppression of the Moiré pattern at the bias of a QW state is in good agreement with our physical model.

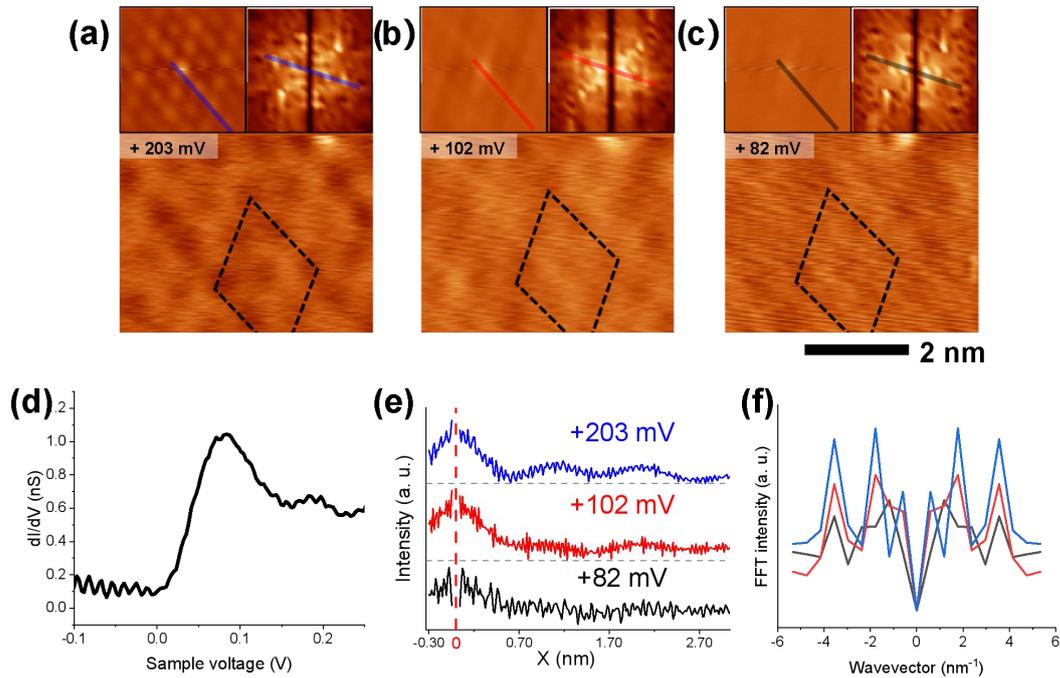

**Fig. SI 9.** dI/dV maps on a 10 nm Al/Ge film. (a), (b) and (c) dI/dV maps taken at bias of +203 mV, +102 mV and +82 mV, respectively. Corresponding self-correlation images and the FFT images are shown in the left and right insets, respectively. (d) dI/dV spectrum taken at the mapping region. A QW peak sits at bias of +79±3 mV. (e) and (f) Intensity Profiles along the solid lines in self-correlation images and FFT images, respectively. These profiles are shown in the corresponding colors of the solid lines.